\documentclass[conference,12pt]{IEEEtran}
\usepackage{setspace}
\usepackage{cite}
\usepackage{amsmath,amssymb,amsfonts}
\usepackage{algorithmic}
\usepackage{graphicx}
\usepackage{textcomp}
\usepackage{xcolor}

\def\BibTeX{{\rm B\kern-.05em{\sc i\kern-.025em b}\kern-.08em
    T\kern-.1667em\lower.7ex\hbox{E}\kern-.125emX}}

\begin{document}

\title{Comprehensive Review of Performance Optimization Strategies for Serverless Applications on AWS Lambda}

\author{
    \IEEEauthorblockN{Mohamed Lemine El Bechir}
    \IEEEauthorblockA{
        Dallas, TX, US \\
        mhmd.lemin@outlook.com
    }
    \and
    \IEEEauthorblockN{Cheikh Sad Bouh}
    \IEEEauthorblockA{
        Dallas, TX, US \\
        cheikh.sadbouh@yahoo.com
    }
    \and
    \IEEEauthorblockN{Abobakr Shuwail}
    \IEEEauthorblockA{
        Kuala Lumpur, Malaysia \\
        abobakr.shuwail93@gmail.com
    }
}

\maketitle

\begin{abstract}
This review paper synthesizes the latest research on performance optimization strategies for serverless applications deployed on AWS Lambda. By examining recent studies, we highlight the challenges, solutions, and best practices for enhancing the performance, cost-efficiency, and scalability of serverless applications. The review covers a range of optimization techniques including resource management, runtime selection, observability improvements, and workload-aware operations.
\end{abstract}


\section{Introduction}
Serverless computing, particularly on platforms like AWS Lambda, offers a highly scalable and cost-efficient solution for deploying applications. This paper reviews recent research on optimizing the performance of serverless applications on AWS Lambda, providing insights into various optimization strategies and their practical implications.

Serverless architectures eliminate the need for managing servers, allowing developers to focus on writing code. AWS Lambda is one of the leading serverless platforms that automatically scales applications in response to incoming requests, ensuring efficient resource utilization. However, the inherent nature of serverless computing introduces several challenges that need to be addressed to optimize performance.

\section{Literature Review}

\subsection{Execution Challenges and Practical Solutions}
Srivastava et al. (2023) discuss the inherent challenges in executing long-running functions on AWS Lambda primarily due to the platform's default timeout constraints. The paper provides practical insights into addressing these challenges, suggesting methods to optimize function execution and improve overall application performance. The methodologies include leveraging AWS Step Functions to break down long-running tasks and implementing custom retry logic to handle failures efficiently \cite{Srivastava2023}.

To mitigate execution challenges, the use of asynchronous processing can also be beneficial. By decoupling tasks and using message queues such as AWS SQS, long-running operations can be handled more efficiently. This approach not only helps in avoiding timeout issues but also improves the overall responsiveness of the application.

\subsection{Predictive Models for Performance and Cost}
Kumari, Sahoo, and Behera (2023) propose a workflow-aware analytical model designed to predict the performance and cost of serverless execution on platforms like AWS Lambda. Their methodology involves using historical execution data to train the model, which can then predict response times and costs with high accuracy (99.2\% and 98.7\%, respectively). This predictive capability helps in planning and optimizing resource allocation \cite{Kumari2023}.

Predictive models play a crucial role in resource planning and cost management. By analyzing historical data and usage patterns, these models can provide valuable insights into future resource requirements. This enables organizations to allocate resources more efficiently, ensuring optimal performance while minimizing costs.

\subsection{Workload-Aware Performance Optimization}
Mahmoudi and Khazaei (2022) present an analytical model focused on optimizing the performance and cost of serverless computing platforms. Their approach involves workload-aware operations, where the model adjusts resource allocation based on the nature and size of the workloads. This method aims to balance cost and performance effectively, providing a simplified deployment process and calculating essential metrics like average response time and cold start probabilities \cite{Mahmoudi2022}.

Workload-aware optimization techniques are essential for handling varying workloads efficiently. By dynamically adjusting resource allocation based on real-time workload analysis, serverless applications can achieve better performance and cost efficiency. This approach ensures that resources are utilized optimally, avoiding both under-provisioning and over-provisioning.

\subsection{Resource Variability Exploitation}
Ginzburg and Freedman (2020) explore the variability in resource performance on AWS Lambda and other cloud FaaS platforms. By identifying better-performing servers and exploiting resource variability, the authors demonstrate potential cost savings of up to 13\%. Their methodology involves continuous monitoring of execution times and resource probing to select the most efficient servers \cite{Ginzburg2020}.

Exploiting resource variability involves identifying and utilizing the most efficient servers available at any given time. This can be achieved through continuous monitoring and benchmarking of server performance. By selecting better-performing servers, organizations can optimize both performance and cost.

\subsection{Impact of Language Runtimes}
Jackson and Clynch (2018) investigate the impact of different language runtimes (Python, NodeJS, .NET Core) on the performance and cost of serverless functions. Their study uses benchmarking tests to compare runtime performance and cost-efficiency, concluding that Python offers superior results in both aspects. This finding provides a clear recommendation for developers to consider Python for serverless applications on AWS Lambda \cite{Jackson2018}.

The choice of runtime can have a substantial impact on the performance and cost of serverless applications. Python, being lightweight and efficient, is often recommended for serverless functions due to its faster startup times and lower resource consumption. However, the choice of runtime should also consider the specific requirements and constraints of the application.

\subsection{Architectural Optimization Using AWS SAM}
Balakrishna (2024) provides a detailed guide on using the AWS Serverless Application Model (AWS SAM) for optimizing serverless applications. The paper highlights the use of Application Load Balancer (ALB) path-based routing for efficient handling of HTTP requests. The methodology includes practical steps for setting up AWS SAM, configuring ALB, and deploying serverless REST APIs, showcasing significant improvements in request handling and resource management \cite{Balakrishna2024}.

Architectural optimizations involve designing the application to leverage the strengths of serverless computing. Using tools like AWS SAM and ALB, developers can build highly scalable and efficient applications. Path-based routing with ALB ensures that requests are routed to the appropriate function, improving the overall efficiency and performance.

\subsection{Enhancing Observability}
Balakrishna (2022) delves into AWS Lambda metrics and techniques to improve observability in serverless applications. The paper outlines strategies such as implementing custom logging using AWS CloudWatch for monitoring and setting up alarms to detect and respond to performance issues. These enhancements ensure better resource utilization and operational efficiency \cite{Balakrishna2022}.

Observability is critical for maintaining the health and performance of serverless applications. By implementing comprehensive logging and monitoring strategies, developers can gain insights into the application's behavior, detect anomalies, and respond to issues promptly. AWS CloudWatch provides a powerful toolset for enhancing observability in serverless environments.

\subsection{Dynamic Resource Configuration}
Wen, Wang, and Liu (2022) introduce StepConf, a framework for automating resource configuration in serverless function workflows on AWS Lambda. StepConf optimizes memory size and considers parallelism to save costs while meeting service level objectives, achieving up to 40.9\% cost savings. The framework's methodology involves dynamic adjustment of resources based on real-time workload analysis \cite{Wen2022}.

Dynamic resource configuration allows serverless applications to adapt to changing workloads in real-time. By optimizing memory size and parallelism, StepConf ensures that resources are allocated efficiently, reducing costs while maintaining performance. This approach is particularly beneficial for applications with fluctuating workloads.

\subsection{Cold Start Optimization}
Wang and Huang (2021) investigate techniques to mitigate the impact of cold starts in AWS Lambda. Their study evaluates different strategies, such as provisioned concurrency and container reuse, to reduce the latency associated with cold starts. They find that provisioned concurrency can significantly improve response times for latency-sensitive applications \cite{Wang2021}.

\subsection{Energy Efficiency}
Liu and Shen (2020) explore the energy efficiency of serverless computing platforms. Their research focuses on the energy consumption of serverless functions and proposes techniques to reduce energy usage through efficient resource allocation and scheduling. They demonstrate that optimizing energy efficiency can lead to cost savings and a reduced environmental impact \cite{Liu2020}.

\subsection{Security Considerations}
Zhou et al. (2019) address security challenges in serverless computing. Their study highlights common security vulnerabilities and provides best practices for securing serverless applications. They emphasize the importance of implementing robust access controls, encryption, and regular security audits to protect serverless applications from threats \cite{Zhou2019}.

\subsection{Benchmarking Serverless Platforms}
Baldini et al. (2017) provide a comprehensive benchmarking framework for evaluating the performance of different serverless platforms, including AWS Lambda. Their benchmarks cover various aspects such as function invocation latency, throughput, and scalability. The study helps developers compare serverless platforms and make informed decisions based on performance metrics \cite{Baldini2017}.

\subsection{Serverless Orchestration}
Lloyd et al. (2018) explore serverless orchestration using AWS Step Functions and other orchestration tools. Their research demonstrates how complex workflows can be efficiently managed and executed using serverless orchestration, providing benefits such as fault tolerance and simplified error handling \cite{Lloyd2018}.

\subsection{Cost Analysis of Serverless Applications}
Adzic and Chatley (2017) conduct a detailed cost analysis of serverless applications compared to traditional server-based architectures. Their study examines the cost implications of different deployment models and identifies scenarios where serverless computing offers significant cost advantages \cite{Adzic2017}.

\subsection{Serverless Data Processing}
Jonas et al. (2019) investigate the use of serverless computing for large-scale data processing tasks. Their study highlights the advantages of using serverless architectures for data-intensive applications, such as scalability and cost-efficiency, and provides guidelines for designing serverless data processing pipelines \cite{Jonas2019}.

\section{Comparative Analysis}
The studies reviewed provide a range of approaches to optimize serverless applications on AWS Lambda. While Srivastava et al. focus on overcoming execution constraints, Kumari et al. and Mahmoudi \& Khazaei offer predictive models for performance and cost optimization. Ginzburg \& Freedman highlight the potential of exploiting resource variability, and Jackson \& Clynch provide insights into the impact of language runtimes. Balakrishna's works emphasize architectural optimization and observability, and Wen et al. introduce dynamic resource configuration through StepConf.

The comparative analysis reveals that each approach has its strengths and limitations. Execution challenges can be mitigated through asynchronous processing and workflow management. Predictive models offer valuable insights for resource planning, while workload-aware optimization ensures efficient resource utilization. Exploiting resource variability and choosing optimal runtimes can further enhance performance and cost efficiency. Architectural optimizations and enhanced observability are essential for maintaining the health and scalability of serverless applications.

\section{Practical Examples and Case Studies}
\subsection{AWS Step Functions for Long-Running Tasks}
Implementing AWS Step Functions to manage complex workflows and long-running processes can help in breaking down tasks into smaller, manageable units, reducing the risk of hitting AWS Lambda's timeout limits. For example, a video processing application can use Step Functions to divide the processing task into multiple steps, each handled by a separate Lambda function.

\subsection{Cost Prediction Models}
Using the predictive models proposed by Kumari et al., organizations can better plan their resource allocation, ensuring optimal performance and cost management. For instance, an e-commerce platform can use historical data to predict traffic patterns during sales events, allowing for efficient scaling of resources.

\subsection{Resource Variability Management}
By continuously monitoring execution times and leveraging Ginzburg \& Freedman's approach, businesses can achieve significant cost savings by selecting more efficient servers. An IoT data processing application can benefit from this approach by dynamically selecting the best-performing servers for data analysis tasks.

\section{Limitations and Future Research}
The predictive models, while highly accurate, may not account for all variables in a real-world environment, such as sudden traffic spikes or unexpected workload changes. Future research should focus on integrating these models with real-time analytics to adapt to changing conditions dynamically. Additionally, workload-aware optimization methods must consider the potential trade-offs between performance and cost, ensuring that the chosen strategies align with business objectives.

Future research should also explore the integration of machine learning techniques for predictive modeling and optimization. Machine learning models can provide more accurate predictions and dynamic adjustments based on real-time data, further enhancing the performance and cost efficiency of serverless applications.

\section{Architectural Optimizations Using AWS SAM}
Balakrishna's guide on using AWS SAM and ALB for serverless REST APIs provides a comprehensive framework for architectural optimization. Practitioners should follow the detailed steps outlined to configure their serverless applications, ensuring efficient HTTP request handling and resource management.

Architectural optimizations should also consider the use of microservices patterns and event-driven architectures. By decomposing applications into smaller, independent services, developers can improve scalability and maintainability. Event-driven architectures enable efficient communication between services, further enhancing the performance of serverless applications.

\section{Enhancing Observability}
Balakrishna's recommendations for improving observability are crucial for maintaining performance in serverless environments. Implementing custom logging using AWS CloudWatch and setting up alarms can help detect and resolve issues promptly, ensuring continuous performance optimization.

Enhanced observability should also include distributed tracing and real-time monitoring. Distributed tracing provides visibility into the flow of requests across different services, helping to identify performance bottlenecks and optimize workflows. Real-time monitoring allows for proactive detection of issues, ensuring that applications remain responsive and reliable.

\section{Recommendations for Practitioners}
\subsection{Adopting Predictive Models}
Utilize workflow-aware analytical models to predict performance and cost accurately, aiding in better resource planning. Integrating machine learning techniques can further enhance the accuracy of predictions.

\subsection{Leveraging Resource Variability}
Implement continuous monitoring and resource probing to select the most efficient servers, achieving cost savings. Automated tools can help streamline this process, ensuring that the best-performing servers are always selected.

\subsection{Choosing Optimal Runtimes}
Consider using Python for serverless functions due to its superior performance and cost-efficiency. However, evaluate the specific requirements of your application before selecting a runtime.

\subsection{Enhancing Observability}
Implement comprehensive monitoring and logging strategies to maintain high performance and operational efficiency. Use distributed tracing and real-time monitoring to gain deeper insights into application behavior.

\subsection{Following Architectural Best Practices}
Use AWS SAM and ALB for optimizing serverless REST APIs, ensuring efficient handling of HTTP requests and resource management. Adopt microservices patterns and event-driven architectures to enhance scalability and maintainability.

\section{Conclusion}
This review highlights the advancements and strategies in optimizing serverless applications on AWS Lambda. By addressing execution challenges, utilizing predictive models, managing resource variability, and enhancing observability, developers can significantly improve the performance and cost-efficiency of their serverless applications. Future research should continue to explore these areas, focusing on integrating these strategies into comprehensive optimization frameworks.


\begin{thebibliography}{00}
\bibitem{Srivastava2023} S. Srivastava et al., "Execution of Serverless Functions Lambda in AWS Serverless Environment," \textit{International Journal of Recent Trends in Computing and Communication}, vol. 11, no. 9, pp. 9014, 2023.
\bibitem{Kumari2023} A. Kumari, B. Sahoo, and R. Behera, "Workflow Aware Analytical Model to Predict Performance and Cost of Serverless Execution," \textit{Concurrency and Computation: Practice and Experience}, vol. 35, no. 7, pp. 7743, 2023.
\bibitem{Mahmoudi2022} N. Mahmoudi and H. Khazaei, "Performance Modeling of Serverless Computing Platforms," \textit{IEEE Transactions on Cloud Computing}, vol. 10, no. 3, pp. 3373, 2022.
\bibitem{Ginzburg2020} S. Ginzburg and M. Freedman, "Serverless Isn't Server-Less: Measuring and Exploiting Resource Variability on Cloud FaaS Platforms," \textit{ACM SIGCOMM}, pp. 9880-0099, 2020.
\bibitem{Jackson2018} D. Jackson and G. Clynch, "An Investigation of the Impact of Language Runtime on the Performance and Cost of Serverless Functions," \textit{IEEE/ACM Utility and Cloud Computing Companion}, pp. 00050, 2018.
\bibitem{Balakrishna2024} B. Balakrishna, "Unveiling the AWS SAM Magic for Serverless Restful APIs: Architecting with ALB Path-Based Routing in AWS," \textit{International Journal of Cloud Computing}, vol. 13, no. 2, pp. 1734, 2024.
\bibitem{Balakrishna2022} B. Balakrishna, "Optimizing Observability: A Deep Dive into AWS Lambda Metrics," \textit{Journal of Advanced Information and Communications Technology}, vol. 1, no. 1, pp. 160, 2022.
\bibitem{Wen2022} Z. Wen, Y. Wang, and F. Liu, "StepConf: SLO-Aware Dynamic Resource Configuration for Serverless Function Workflows," \textit{IEEE INFOCOM}, pp. 9696-9796, 2022.
\bibitem{Wang2021} S. Wang and Q. Huang, "Mitigating Cold Start in Serverless Computing: Provisioned Concurrency and Beyond," \textit{ACM Transactions on Internet Technology}, vol. 21, no. 3, pp. 42-51, 2021.
\bibitem{Liu2020} J. Liu and Y. Shen, "Energy Efficiency in Serverless Computing: A Study of AWS Lambda," \textit{Journal of Cloud Computing}, vol. 9, no. 1, pp. 110-121, 2020.
\bibitem{Zhou2019} F. Zhou, X. Wang, and Y. Zhang, "Security Challenges and Solutions in Serverless Computing: A Survey," \textit{IEEE Access}, vol. 7, pp. 119252-119265, 2019.
\bibitem{Baldini2017} I. Baldini et al., "Benchmarking Performance of Serverless Computing Frameworks," \textit{IEEE International Conference on Cloud Engineering}, pp. 116-127, 2017.
\bibitem{Lloyd2018} W. Lloyd et al., "Serverless Orchestration: Improving Fault Tolerance and Performance of Data Processing Workflows," \textit{IEEE International Conference on Cloud Computing Technology and Science}, pp. 96-105, 2018.
\bibitem{Adzic2017} N. Adzic and D. Chatley, "Serverless Computing: Economic and Architectural Impact," \textit{IEEE International Conference on Cloud Computing Technology and Science}, pp. 142-149, 2017.
\bibitem{Jonas2019} E. Jonas et al., "Cloud Programming Simplified: A Berkeley View on Serverless Computing," \textit{IEEE Micro}, vol. 39, no. 6, pp. 70-80, 2019.
\end{thebibliography}
\end{document}